\documentclass[runningheads]{llncs}
\usepackage{graphicx}
\usepackage[T1]{fontenc}   
\usepackage{txfonts}
\usepackage{mathptmx}
\usepackage[pdflang={en-US},pdftex]{hyperref}
\usepackage{color}
\usepackage{booktabs}
\usepackage{textcomp}
\usepackage{paralist}
\usepackage[numbers]{natbib}  
\usepackage[nolist]{acronym}

\newcommand{\quotes}[1]{``#1''}

\begin{document}
\title{Making Things Explainable vs Explaining: Requirements and Challenges under the GDPR}
%
%
\author{Francesco Sovrano\inst{1}\orcidID{0000-0002-6285-1041} \and
Fabio Vitali\inst{1}\orcidID{0000-0002-7562-5203} \and
Monica Palmirani\inst{2}\orcidID{0000-0002-8557-8084}}
\authorrunning{F. Sovrano et al.}
\titlerunning{The Difference between Making Things Explainable and Explaining}
%
\institute{DISI, University of Bologna \and
CIRSFID-AI, University of Bologna}
\maketitle              
\begin{abstract}
    The \acf{EU} through the \acf{AI-HLEG} and the \acf{GDPR} has recently posed an interesting challenge to the \acf{XAI} community, by demanding a more user-centred approach to explain \acf{ADMs}.
	Looking at the relevant literature, \ac{XAI} is currently focused on producing explainable software and explanations that generally follow an approach we could term \textit{One-Size-Fits-All}, that is unable to meet a requirement of centring on user needs.
	One of the causes of this limit is the belief that \textit{making things explainable} alone is enough to have \textit{pragmatic explanations}.
	Thus, insisting on a clear separation between \textit{explainabilty} (something that can be explained) and \textit{explanations}, we point to \acf{YAI} as an alternative and more powerful approach to win the \ac{AI-HLEG} challenge.
	\ac{YAI} builds over \ac{XAI} with the goal to collect and organize explainable information, articulating it into something we called user-centred explanatory discourses.
	Through the use of explanatory discourses/narratives we represent the problem of generating explanations for \acp{ADM} into the identification of an appropriate path over an explanatory space, allowing explainees to interactively explore it and produce the explanation best suited to their needs.
\keywords{Trustworthy AI \and explanatorY AI (YAI) \and XAI \and HCI}
\end{abstract}
	\section{Introduction}
	The academic interest in \ac{AI} \cite{eucommission18} has grown together with the attention of Countries and people towards the possibly disruptive effects of \ac{ADM} \cite{eucommission16} in industry and the public administration (e.g., COMPAS \cite{flores2016false}, or in Italy the case-law "Buona Scuola"\footnote{Cons. stato, sez. VI, sent. 8 aprile 2019, n. 2270, Cons. Stato, sez. VI, sent. del 13 dicembre 2019, n. 8472, Cons. Stato, sez. VI, sent. del 4 febbraio 2020, n. 881.}), effects that may affect the lives of billions of persons \cite{kuhn2019application}. 
	Therefore governments are starting to act towards the establishment of ground rules of behaviour from complex systems, for instance through the enactment of the European \ac{GDPR}\footnote{Regulation (EU) 2016/679.}, which identifies \textit{fairness}, \textit{lawfulness}, and in particular \textit{transparency} as basic principles for every data processing tools handling personal data; even identifying a new \textit{right to explanation} for individuals whose legal status is affected by a solely-automated decision. 
	As a result, several expert groups, including those acting for the European Commission, have started asking the \ac{AI} industry to adopt ethics code of conducts as quickly as possible \cite{cath2018artificial,floridi2018ai4people}, drawing a set of expectations to meet in order to guarantee a \textit{right to explanation}. These expectations define the goal of explanations under the \ac{GDPR} and thus describe the requirements for explanatory content.
	Many interpretations have been given of what qualifies an explanation in this context, but among them we mention the one by the \ac{AI-HLEG}, for its relevance and prominence.
	The \ac{AI-HLEG} was established in 2018, by the European Commission, with the explicit purpose of applying the principles of the \ac{GDPR} specifically to \ac{AI} software, and produced a list of fundamental ethical principles for \emph{Trustworthy \ac{AI}} tools that include \emph{fairness} and \emph{explicability}. The \emph{explicability} principle, in particular, means to provide alternative measures in case of \quotes{black box} algorithms like \quotes{traceability, auditability and transparent communication on system capabilities}, in order to respect the fundamental rights. So it is important to provide information about \emph{how} the \ac{ADM} works, \emph{what} is the final decision, \emph{why} the \ac{ADM} provides such conclusion, \emph{which} data are used for training the AI and for the concrete real case processing. \emph{Explicability} concerns the \emph{ex-post} processing but also the \emph{ex-ante} informative communication.
	Most importantly, according to the \ac{AI-HLEG}, explanations should be \quotes{adapted to the expertise of the stakeholder concerned (e.g. layperson, regulator or researcher)} and more over it \quotes{highly dependent on the context} \cite{hleg2019ethics}, putting individual's needs at the centre, in a challenging way.
	 
	Notwithstanding these quite recent efforts, understanding what constitutes an explanation is a long-standing open problem. In literature there are various efforts in this direction and a long history of debates and philosophical traditions, often rooted in Aristotle's works and those of other philosophers. Among the many models proposed over the last few centuries some are now considered fallacious, albeit historically useful (e.g. \citeauthor{hempel1965aspects}'s one \cite{hempel1965aspects}), in favour of more pragmatic (user-centred) ones (e.g. \citeauthor{achinstein2010evidence}'s \cite{achinstein2010evidence}).
	Despite this, \citeauthor{hempel1965aspects}'s theory and \citeauthor{salmon1984scientific}'s \textit{Causal Realism} are probably the most (implicitly) mentioned and adopted models for explanations in \ac{AI}, raising the question of whether technology is really aligned to the understandings of regulators and society or it is just acting conveniently.
	In fact, most of the literature on \ac{AI} and explanations (e.g. eXplainable AI \cite{arrieta2020explainable}) is currently focused on one-size-fits-all approaches usually able to produce only one type of explanations, defined through causal lens. Additional literature is focused on argumentation theory \cite{Cocarascu2020argumentation} or on sub-symbolic methodologies \cite{Calegari2020subsymbolic} for providing a deductive or inductive explanation.
	
	It appears that this focus on pursuing one-size-fits-all explanations in \ac{XAI} is justified by convenient definitions framing an explanation as the product of an act of making things explainable rather than a pragmatic (user-centred) act of explaining based on explainability.
	In other terms, there is no clear distinction between \textit{making things explainable} and actually \textit{explaining}. 
	The exceptions to this pattern seem to be still too rare to be representative of disciplines like \ac{XAI}.
	In this paper we take a strong stand against the idea that static, one-size-fits-all approaches to explanation have a chance of being pragmatic, thus meeting the \ac{AI-HLEG} guidelines, and we propose to adopt a strong logical separation between \textit{explainability} and \textit{explaining}.
	In fact, we argue that explaining to humans is \textit{computationally irreducible} and one-size-fits-all approaches (in the most generic scenario) may suffer the curse of dimensionality as soon as the complexity of the explanandum surpasses a fairly trivial threshold. 
	For example, a complex big-enough \textit{explainable} software can be super hard to \textit{explain}, even to an expert, and the optimal (or even sufficient) explanation might change from expert to expert. In this specific example, an explainable software is necessary but not sufficient for explaining.
	This is why we first draw a clear separation between \ac{XAI} and \acf{YAI}, which refers to systems that (given a \quotes{traditional} \ac{XAI} system) are actually able to produce a satisfactory explanation ready to be delivered to a human user interested in examining the complex working and output of the system. 
	Subsequently, we propose a model for \ac{YAI} shaped on \textit{discursive explanations}. Discursive explanations give a strong background of principles and means to create an interactive explanatory system that is able to produce user-centred explanations, by providing an explanatory space that is amenable to exploration by the users in order to create the explanation that best suits each one's background, needs and objectives.
	
	This paper is structured as follows. 
	In Section \ref{sec:background} we provide an introduction to the \ac{GDPR} and the \textit{Right to Explanation}, and we also provide a brief summary of the \ac{AI-HLEG} Guidelines for Trustworthy \ac{AI}.
	In Section \ref{sec:problem_statement}, taking off from the \ac{GDPR} and the \ac{AI-HLEG} guidelines, we give a motivation of why user-centred explanatory tools are a key ingredient for Trustworthy \ac{AI}. In this section we discuss the most prominent \ac{XAI} issues to this end and the problem of \textit{computational irreducibility} in explanations.
	In Section \ref{sec:proposed_solution} we give an high-level overview of a possible model of User-Centred Explanatory Tool, defining \ac{YAI} as a Explanatory Discursive Process responsible to collect and structure explainable information articulating it into user-centred explanations. Finally, in section \ref{sec:conclusions} we conclude with a brief recap, pointing to a proof of concept.
	
	
	\section{Background: the Right to Explanation} \label{sec:background}
	The \acf{GDPR} is an important 2016 EU regulation on personal data protection and the connected freedoms and rights. Since the \ac{GDPR} is technology-neutral, it does not directly refer to \ac{AI}, but several provisions are highly relevant to the use of \ac{AI} for \acf{ADM}.
	For instance \cite{ico2019}:
	\begin{itemize}
		\item Principle 1. (a) requires personal data processing to be fair, lawful, transparent, necessary and proportional (Articles 5). 
		\item Article 12 defines the obligations to fulfil a transparent information, communication and the modalities for the exercise of the data subject's rights.
		\item Articles 13-14-15 give individuals the right to be informed of the existence of solely automated decision-making, meaningful information about the logic involved, and the significance and envisaged consequences for the individual. 
		\item Article 22 gives individuals the right not to be subject to a solely automated decision producing legal or similarly significant effects. 
		\item Article 22(3) obliges organizations to adopt suitable measures to safeguard individuals when using solely automated decisions, including the right to obtain human intervention, to express his or her view, and to contest the decision. 
	\end{itemize}
	Art. 22 defines the right to claim of a human intervention when a completely \acf{ADMs} may affect the legal status of a citizen. Art. 22 includes also several exceptions that derogate \quotes{to be subject to a decision based solely on automated processing} when the legal basis are supported by contract, consent or law.
	These conditions significantly limit the potential applicability of the right to explanation. For this reason in case of contract or consent the art. 22, paragraph 3 introduces the \quotes {right to obtain human intervention on the part of the controller, to express his or her point of view and to contest the decision}. 
	Here explanations seem to be provided only after decisions have been made (\emph{ex-post} explanations), and are not a required precondition to protest decisions. This is not completely true: in arts. 13-14-15 there is the obligation to inform about the \quotes {the existence of automated decision-making, including profiling, referred to in Article 22(1) and (4) and, at least in those cases, meaningful information about the logic involved (Recital 63), as well as the significance and the envisaged consequences of such processing for the data subject.} (\emph{ex-ante} explanations). This combination of articles make the right of explanation very articulated and composed of different stages. 
	Additionally, the recent White Paper on Artificial Intelligence \cite{comm202065} emitted by the European Commission stressed the need to monitor and audit not only the \acf{ADM} algorithms but also the data records used for training, developing, running, the \ac{AI} systems in order to fight the opacity and to improve transparency. 
	From a technical point of view, there are technology-specific information to consider in order to fully meet the explanation requirements of the \ac{GDPR}, for a more detailed overview refer to \cite{sovrano2020modelling}.
	The qualities of explanations are listed in different works \cite{mueller2019explanation}, but the EU Parliament \cite{euparliament2019} lists the following as a good summary of the current state of the art: intelligibility, understandability, fidelity, accuracy, precision, level of detail, completeness, consistency.
	
	Article 22 is open to several interpretations \cite{wachter2017right,pagallo2020,palmirani2020} about whether providing individualised explanations is mandatory or just a good practice. To this end, Recital 71 provides interpretative guidance of Article 22. 
	Two items are missing in Article 22 relative to Recital 71: the provision of \quotes{specific information} and the \quotes{right to obtain an explanation of the decision reached after such assessment}. The second omission in particular raises the issue of whether controllers are really required by law to provide an individualised explanation.
	This issue is partially tackled by the \ac{AI-HLEG} guidelines (endorsed by the EU Commission), giving further reason to believe that there is the intention to prefer user-centred explanations as soon as the technology is mature enough to guarantee them. At contrary Recital 63 requires \emph{ex-ante} that the data subject should  have the right to know and obtain communication in particular with regard to \quotes{the logic involved in any automatic personal data processing}.
	The \ac{AI-HLEG} tries to extend the \ac{GDPR} expectations, targeting \ac{AI} and giving further guidelines: accessibility and universal design should be a requirement for Trustworthy \ac{AI}, with user-centrality at the core. 
	This idea of a user-centred explanatory process find its roots in philosophy, for example in:
	\begin{itemize}
		\item Ordinary Language Philosophy \cite{achinstein1983nature,mayes2005theories}: the act of explanation as the illocutionary attempt to produce understanding in another by answering questions in a pragmatic way.
		\item Cognitive Science \cite{holland1989induction,mayes2005theories}: explaining as a process of belief revision, etc..
	\end{itemize}
	
	
	\section{Problem Statement} \label{sec:problem_statement}
	Some of the limits in the current generation of \ac{XAI} approaches have already been identified and spelt out by existing literature: 
	\begin{itemize}
		\item \quotes{\ac{XAI} has produced algorithms to generate explanations as short rules, attribution or influence scores, prototype examples, partial dependence plots, etc. However, little justification is provided for choosing different explanation types or representations} \cite{wang2019designing}. 
		\item \quotes{Research on explanation is typically focused on the person (or system) producing the explanation. [\dots]  Does the explainee understand the system, concepts, or knowledge?} \cite{mueller2019explanation}.
		\item \quotes{Much of \ac{XAI} research tended to use the researchers' intuition of what constitutes a good explanation. There exist vast and valuable bodies of research in philosophy, psychology, and cognitive science of how people define, generate, select, evaluate, and present explanations, which argues that people employ certain cognitive biases and social expectations to the explanation process.} \cite{miller2018explanation}
		\item \quotes{\ac{XAI} systems are built for developers, not users.} \cite{miller2017explainable,mueller2019explanation}
		\item etc..
	\end{itemize}
	To summarize, despite several efforts (e.g. \cite{miller2018explanation,darpa2016broad}) to tackle these issues, we can notice a majority of \ac{XAI} tools lacking:
	\begin{enumerate}
		\item A broader vision: \ac{XAI} should not involve only computer science, but also philosophy, psychology, cognitive science, etc..
		\item Focus on user-centrality. 
		\item A consistent approach to evaluate the quality of explanations.
	\end{enumerate}
	We claim that the cause of these limits are in the misunderstanding that explainability is enough for explaining.
	Indeed, by insisting on a clear logical separation between explainable systems and actual explanations, we argue that \ac{XAI} is necessary but not sufficient for Trustworthy \ac{AI}. 
	In fact, \ac{XAI} seems to be currently focused on producing explainable software and explanations that generally follow only a One-Size-Fits-All approach, failing to meet the user-centrality requirements.
	In the most generic scenario, explanations following a One-Size-Fits-All approach (\emph{OSFA explanations}) should be considered not user-centred, by construction. For example, static representations where all aspects of a fairly long and complex computation are described and explained are one-size-fits-all explanations. 
	
	OSFA explanations have intuitively at least two problems:
	\begin{enumerate}
		\item if they are small enough to be simple, then in a complex enough domain they would not be able to generate an explanation containing enough information to satisfy the explanation appetite of every user, as the quantity of details required for satisfying every user would be necessarily larger than any small explanation in a few words.
		\item if they contain all the necessary information, in a complex-enough domain they would contain an enormous amount of content and users interested in a specific aspect of the explanation would need to look for it within the whole explanation in hundreds or thousands of explanatory items mostly irrelevant to their purposes.
	\end{enumerate}	
	OSFA explanations could be useful for simple domains, but the complexity of a domain is exactly what motivates the need for explanations. In other terms, usefulness of explanations is obviously greater in complex domains.
	
	An interesting parallel, to show the second problem, is that of surveillance cameras in front of a bank door.
	Surveillance cameras continuously record and make available to the investigators hundreds and hundreds of hours of excellent quality videos that allow the precise identification of thousands of people passing under the cameras. 
	But our investigator is not interested in hundreds of hours of video, but only in those three seconds in which a suspect person in need to be identified was under the cameras. 
	The relevance of these few seconds (out of hundreds of hours) is entirely based on the specific investigative task, which depends on the function that the investigator gives to the identification of the person, and this function depends on the purpose of identification (i.e. Is he the robber? A possible accomplice? A witness?). 
	The purpose of the investigation is known to the investigator but not to the surveillance system, and in many cases it cannot be decided in advance but it becomes clear only during the evolution of the investigation. 
	Similarly, the interest of a user in the output of an explanation system often may lie on a few short statements out of the hundreds of thousands that the explanation system may be able to generate, and these few ones depend on the function that the user gives to the explanation. This is why we must assume that in general the purpose of the explanation is known to the user but not to the explanation system, and it cannot be decided in advance but it becomes clear only during the evolution of the task in which the explanation is required. 
	This phenomenon is known also as \textit{computational irreducibility} \cite{zwirn2013unpredictability} and it is typical of emerging phenomena, such as physical, biological and social ones \cite{beckage2013more}.
	
	A user-centred explanatory tool requires to provide goal-oriented explanations. Goal-oriented explanations implies explaining facts that are relevant to the user, according to her/his background knowledge, interests and other peculiarities that make her/him a unique entity with unique needs that may change over time.
	The computational irreducibility issue raises the following questions:
	\begin{enumerate}
		\item How to model and create a \textit{user-centred} explanatory process, without rewriting the tool for every different user?
		\item How to evaluate the quality of an explanatory process?
	\end{enumerate}

	\section{Proposed Solution} \label{sec:proposed_solution}
\begin{figure}
		\centering
		\includegraphics[width=.9\columnwidth]{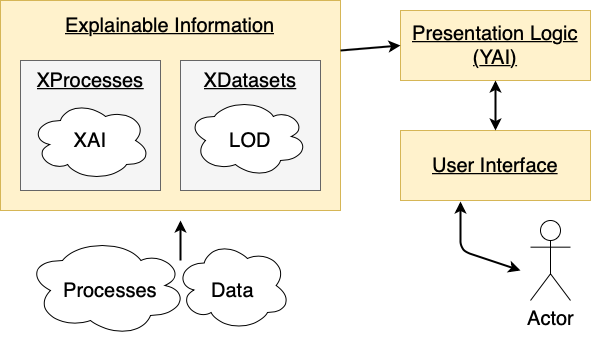}
		\caption{XAI vs YAI: an abstract model of Explanatory Tool for Trustworthy AI. This model shows how to decompose the flow of explanatory information that moves from raw representations of processes/data to the explainee (or actor). Raw data are refined into explainable datasets - e.g. Linked Open Data (LOD), etc.. Raw processes are refined into explainable processes. Explainable information can be used by YAI to generate pragmatic explanations.}
		\label{fig:xai_vs_yai}
	\end{figure}

	In order to answer the first question we propose to:
	\begin{itemize}
		\item Disentangle \textit{explainability} from \textit{explaining}: that is separate the presentation logic (\textit{explaining}) from the application logic (\textit{explainability}). In fact, only \textit{explaining} has to be user-centred.
		\item Design a presentation logic that would allow personalised explanations given the same explainable information.
	\end{itemize}
	In figure \ref{fig:xai_vs_yai} we show a simple model of an Explanatory Tool for Trustworthy AI, obtained by our own need to clearly separate between explainabilty and explanations.
	More in detail, to increase the overall cohesion of the system, in this model we require an explicit logical separation between the functionalities related to \emph{producing explainable information}, and those related to \emph{producing pragmatic explanations}. In addition, we envision another logical separation in the production of actual explanations between \emph{building explanations} (i.e. the presentation logic) and \emph{interfacing with users}. Independently, producing explainable information should be separated in \emph{generating explainable processes} and \emph{producing explainable data-sets}.
	Thus, the main modules involved in the model are:
	\begin{itemize}
		\item The \acf{EI} module, made of the \acf{XP} and the \acf{XD} sub-modules.
		\item The \ac{YAI} or Presentation Logic module.
		\item The \acf{UI} module.
	\end{itemize}
	In other terms, we propose to distinguish between \acf{XAI} and \acf{YAI}, considering them as different components of Trustworthy \ac{AI}. We like to say that Trustworthy \ac{AI} needs both the Xs and the Ys of \ac{AI}\footnote{XX and XY are human chromosomes responsible for gender.}.
	
	The \ac{YAI} module is the module responsible to collect and structure explainable information articulating it into user-centred explanations. In other terms, defining the \ac{YAI} module is the same of defining a \textit{user-centred explanatory process}. 
	We are interested in defining a user-centred explanatory process aligned to the \ac{GDPR} and the \ac{AI-HLEG} guidelines.
	Speaking of user-centrality, we may assume that different types/groups of users exist: lay person, expert, legal operators, etc.. each one with its own background knowledge and unique characteristics.
	If the explanations have to be tailored, does this imply that we should have a different explanatory tool for every possible different user? Probably not.
	We believe that an explanatory tool is an instrument for articulating explainable information into an \textit{explanatory discourse}.
	This definition of explanatory tool is drawn from the essential best-practices of scientific inquiry, involving \cite{berland2009making}: 
	\begin{itemize}
		\item Sense-making of phenomena: classical question answering to collect enough information for understanding, thus building an explainable explanandum (perhaps through \ac{XAI}).
		\item Articulating understandings into discourses: re-ordering and aggregation of explainable information to form an explanatory narrative or more generally a discourse to answer research questions.
		\item Evaluating: pose and answer questions about the quality of the presented information; e.g., argument them in a public debate.
	\end{itemize}
	Therefore we define a user-centred explanatory discourse as: \quotes{A sequence of information (explanans) to increase understanding over explainable data and processes (explanandum), for the satisfaction of a specified explainee that efficiently and effectively interacts with the explanandum (interaction) having specific goals in a specified context of use}.
	Our definition takes inspiration from \cite{passmore1962explanation,norris2005theoretical,lipton2001good}, integrating concepts of usability defined in ISO 9241 (Ergonomics of Human System Interaction \cite{international2010ergonomics}), such as the insistence on the term \quotes{specific}, the triad \quotes{explainee}, \quotes{goal} and \quotes{context of use}, as much as the identification of specific quality metrics, which in our case are \textit{effectiveness}, \textit{efficiency} and \textit{satisfaction}. 
	
	Similarly to how \emph{satisfaction} has increased in importance in user experience studies in recent years, we believe that satisfaction should be considered one of the most important metrics for the assessment of the quality of explanations, too. The qualities of the explanation that provides the explainee with the necessary \textit{satisfaction}, using the categories provided by \cite{norris2005theoretical}, can be summarized in a good choice of narrative appetite, structure and purpose. To understand \quotes{narrative appetite} we have to consider that \quotes{in order for a narrative discourse to flourish, both parties (the narrator and the reader) have to find engagement in this social transaction interesting enough to prevail over competing activities. Thus, stories must not only be accounts of events, but accounts of events that someone cares to know more about; we must want to know what happened if we are to continue reading or listening.} This appetite can be quenched by the proper structuring of the narration: \quotes{Narrative, we have shown, is a narrator’s recounting of \textit{events structured in time}. The elements of both time and structure are associated in many descriptions of narrative}. In addition, \quotes{The element of \textit{connectability} [\dots] structures different texts. Connectability [\dots] must be strictly observed in expository texts where an argument is to be developed or information is to be conveyed. In such texts, the writer aims for a precise interpretation where a multiplicity of possible meanings must be constantly narrowed down}. Finally, the identification of purpose in narratives is central: \quotes{stories are constructed to help us understand the world we live in: to help comprehend the life that is in me and around me. [\dots] it is through narrative that we are able to accommodate the new within that which is familiar to us. In these descriptions of purpose, narrative can be interpreted as helping us better understand the natural as well as the human world}.
	
	The problems of a user-centred approach to explanations is that fully-automated explanatory processes are unlikely to target quality parameters that guarantee the satisfaction of all specified explainees, as described above, due to the \textit{computational irreducibility} of the process of explaining.  
	Even if an \ac{AI} could be used to generate such user-centred explanations, in the context of explanations under the \ac{GDPR} this would only shift the problem of explaining from the original \ac{ADM} to another \ac{ADM} (the explanatory \ac{AI} that explains the original \ac{ADM}).
	As such we believe that (at least for the explanations under the \ac{GDPR}), the most straightforward solution is to encourage readers (explainees) and narrators (explainers) to become one, \emph{users} generating the narration for themselves by selecting and organizing narratives of individual event-tokens according to the structure that best caters their appetite and purpose. 
	In this sense, a tool for creating explanatory discourses would allow users to build intelligible sequences of information, containing arguments that support or attack the claims underlying the goal of an \textit{explanatory narrative process}.
	This idea of data controllers and data subjects \quotes{becoming one} can be understood in a twofold way. First, at its best possible light, such tool should convince and dissuade data subjects to ask for human intervention, e.g. Art 22(3) of the GDPR. Second, the tool should help data controllers to abide by the law, by illustrating the decision that can be contested by data subjects. 
	
	An \textit{explanatory narrative} is always only one of the many possible narratives that can be built to shed light on an explanandum. All the possible narratives for an explanandum form a complex network of information that we call \textit{Explanatory Space}. In this sense, an explanatory discourse is a path within an Explanatory Space.
	As analogy, we might see the Explanatory Space as a sort of manifold space where every point within it is interconnected information about one or more aspects of the explanandum. So that every point of the Explanatory Space is not user-centred locally, but globally as an element of a sequence of information that can be chosen by a user according to its interest drift while exploring the space.
	
	As mentioned in Section \ref{sec:problem_statement}, the amount of information forming such Explanatory Spaces can be overwhelming, given any complex-enough explanandum. Thus, in order to answer our research question, what we need is to design a process to effectively allow users to extract explanatory narratives from an Explanatory Space.
	In \cite{sovrano2020modelling} we present our model of Explanatory Narrative Process making specific references to the \ac{GDPR} and the \ac{AI-HLEG} guidelines, modelling a generic explanatory process, giving a formal definition of  explanandum, explanans and Explanatory Space. Hereafter we show a plausible example of \ac{YAI} in action.
	
	\subsection{Example} \label{sec:example}
\begin{figure}
		\includegraphics[width=1.\columnwidth]{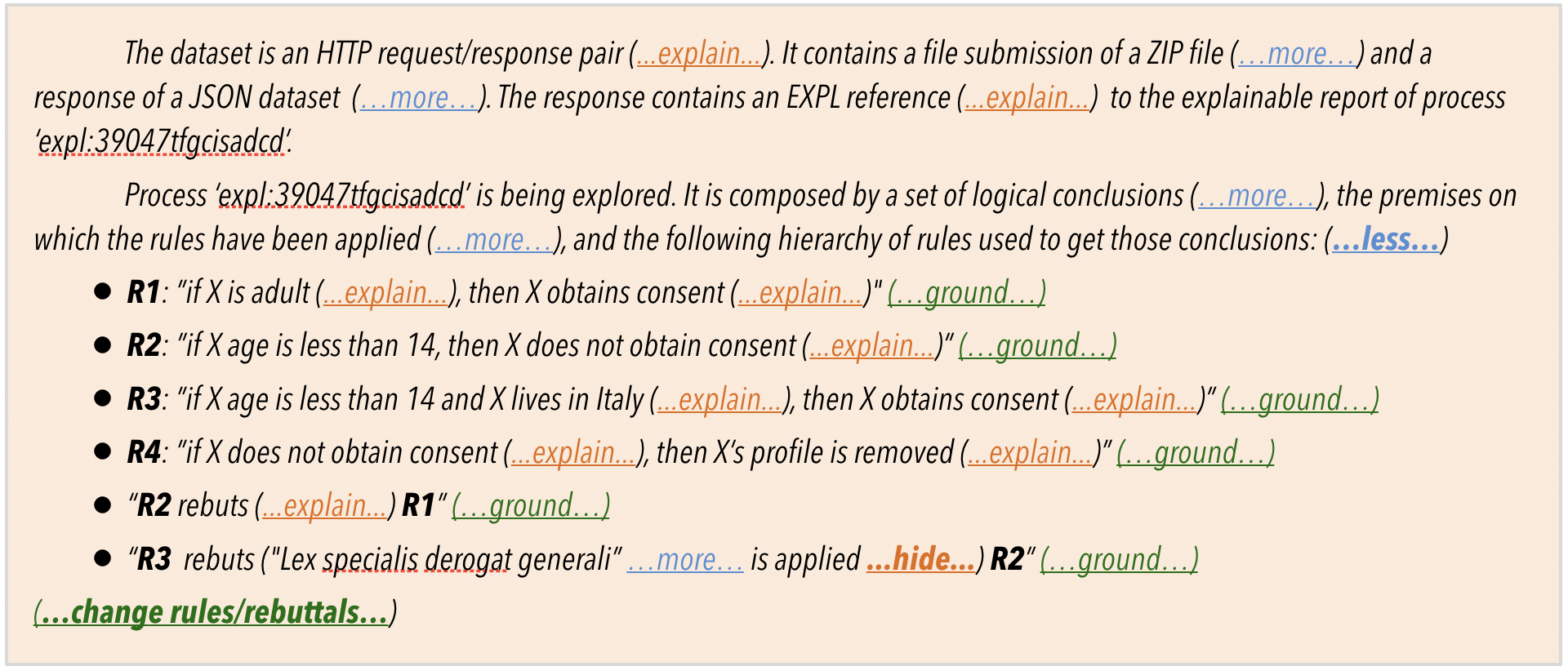}
		\caption{Example of explainer: underlined coloured words represent different possible actions a user can operate to explore the Explanatory Space, extracting its own narrative. For example, clicking on a \quotes{...more...} button the user can expand the explanans.}
		\label{fig:screen3}
	\end{figure}

	Let's consider the following example where a user-centred explanatory tool is used to explain the decision taken by an \ac{ADM} on a case concerning the \ac{GDPR}, art. 8.
	The aforementioned case is about the conditions applicable to child's consent in relation to information society services. The art. 8 of \ac{GDPR} fixes at 16 years old the maximum age for giving the consent without the parent-holder authorization. This limit could be derogated by the domestic law. In Italy the legislative decree 101/2018 defines this limit at 14 years. In this situation we could model legal rules in LegalRuleML \cite{athan2013oasis,palmirani2018modelling} using defeasible logic,
	in order to be able to represent that the \ac{GDPR} art. 8 rule (16 yearsOld) is overridden by the Italian's (14 yearsOld). 
	The SPINDle legal reasoner processes the correct rule according to the jurisdiction (e.g., Italy) and the age. 	
	Suppose that Marco (a 14 years old Italian teenager living in Italy) uses Whatsapp, and his father, Giulio, wants to remove Marco's subscription to Whatsapp because he is worried about the privacy of Marco when online. 
	In this simple scenario, the \acf{ADM} system would reject Giulio's request to remove Marco's profile, because of the Italian legislative decree 101/2018.
	What if Giulio wants to know the reasons why his request was rejected?
	Figure \ref{fig:screen3} shows a possible view of a user-centred explanatory tool based on our model. Thanks to the user-centred explanatory tool Giulio can actually choose what information to expand and consider, building its own personalised explanatory discourse out of a predefined Explanatory Space.
	
	\section{Conclusions} \label{sec:conclusions}
	In this paper we analysed some of the limits in the current generation of \ac{XAI} approaches, with respect to the goals of Trustworthy \ac{AI} set by the \ac{GDPR} and the \ac{AI-HLEG} guidelines, identifying the cause of these limits in the misunderstanding that \textit{making things explainable} is enough for \textit{pragmatically explaining}.
	Indeed, by insisting on a clear logical separation between explainable systems and actual explanations, we argued that \ac{XAI} is necessary but not sufficient for Trustworthy \ac{AI}, therefore presenting an abstract model of \acf{YAI}. 
	In our model, \ac{YAI} builds over \ac{XAI} and it is intended to be a set of tools for organising the presentation logic of a user-centred explanatory software in a way that would allow personalised explanations about complex-enough explananda by generating \textit{discursive explanations} out of an Explanatory Space.
	In this paper we take a strong stand against the idea that static, one-size-fits-all approaches to explanation have a chance of being pragmatic, thus meeting the \ac{AI-HLEG} guidelines.
	For a concrete proof of concept of \ac{YAI} (including software and experiment analysis) we point the reader to our most recent works, e.g. \cite{sovrano2021philosophy}.
	
	Finally, it is clear that the solution we proposed avoids the problem which relates to balancing between what is possible in terms of formal explainability and what is required as to the level of detail of information regarding the \quotes{logic of processing}. In other words, we assumed that systems in question can be both formally explainable and pragmatically able to be explained. 
	So, we leave as future work an analysis of what are the minimum requirements for information to be considered explainable enough for pragmatic explanations with a proper degree of exactness, detail and fruitfulness\footnote{Perhaps drawing from Carnap's theory \cite{novaes2017carnapian}.}. This might help also to perform a reasonable impact assessment of the \ac{ADM}, as defined by art. 35 of the \ac{GDPR}.
	
	\section*{Acknowledgements}
	This work was partially supported by the European Union's Horizon 2020 research and innovation programme under the MSCA grant agreement No 690974 \quotes{MIREL: MIning and REasoning with Legal texts}. Last but not least, a big thank you to all the reviewers for their brilliant comments and different insights.

\bibliographystyle{plainnat}
\bibliography{biblio}

\begin{acronym}
	\acro{EU}{European Union}
	\acro{ADM}{Automated Decision-Making system}
	\acro{ADMs}{Automated Decision-Making systems}
	\acro{AI-HLEG}{High-Level Expert Group on Artificial Intelligence}
	\acro{AI}{Artificial Intelligence}
	\acro{XAI}{eXplainable AI}
	\acro{YAI}{explanatorY AI}
	\acro{HCI}{Human-Computer Interaction}
	\acro{RL}{Reinforcement Learning}
	\acro{EN}{Explanatory Narrative}
	\acro{ENs}{Explanatory Narratives}
	\acro{EP}{Explanatory Process}
	\acro{ES}{Explanatory Space}
	\acro{GDPR}{General Data Protection Regulation}
	\acro{ETTAI}{Explanatory Tool for Trustworthy AI}
	\acro{EI}{Explainable Information}
	\acro{XP}{eXplainable Processes}
	\acro{XD}{eXplainable Datasets}
	\acro{UI}{User Interface}
	\acro{RDF}{Resource Description Framework}
	\acro{AIX360}{AI Explainability 360}
	\acro{CEM}{Contrastive Explanations Method}
	\acro{UCET}{User-Centred Explanatory Tool}
	\acro{SET}{Static Explanatory Tool}
	\acro{KB}{Knowledge Base}
	\acro{TFIDF}{Term Frequency–Inverse Document Frequency}
	\acro{USE}{Universal Sentence Encoder}
	\acro{SUS}{System Usability Scale}
\end{acronym}
\end{document}